\documentclass[preprint,12pt,authoryear]{elsarticle}

\usepackage{amssymb}
\usepackage{amsmath}

\journal{Nuclear Physics B}

\usepackage{amssymb}
\usepackage{cite}
\usepackage{amsmath,amssymb}
\usepackage{algorithmic}
\usepackage{xcolor}
\usepackage[ruled,linesnumbered,lined]{algorithm2e}
\usepackage[normalem]{ulem}

\usepackage{hyperref}
\usepackage{footnote}
\makesavenoteenv{table}
\makesavenoteenv{tabular}
\usepackage{color}
\usepackage{tablefootnote}
\usepackage{nomencl}
\makenomenclature

\usepackage{lipsum}
\usepackage{mathtools}
\usepackage{cuted}
\usepackage{graphicx}
\usepackage{subfig}
\usepackage{multirow}
\usepackage{array}
\usepackage{comment} 
\usepackage{amsmath}
\newlength\myindent
\begin{document}

\begin{frontmatter}



\title{An Accurate Optimal Power Flow for AC-DC Distribution Network with Distributed Energy Resources}

\author{Md Shamim Hasan\textsuperscript{1*}, Md Mahmud-Ul-Tarik Chowdhury\textsuperscript{2}, Krishna Murari\textsuperscript{3}, and Sukumar Kamalasadan\textsuperscript{4}} 

\affiliation{organization={$^{1}$Old Dominion University,},
            addressline={5115 Hampton Road Blvd}, 
            city={Norfolk},
            postcode={23529}, 
            state={VA},
            country={USA} }
            
\affiliation{organization={$^{1}$Dominion Energy,},
            addressline={600 E. Canal St.}, 
            city={Richmond},
            postcode={23219}, 
            state={VA},
            country={USA}}
             \vspace{-3mm}
\affiliation{organization={$^{3}$Indian Institute of Technology Bhilai,},
            addressline={Dept.of Electrical Engineering}, 
            city={Bhilai},
            postcode={}, 
            state={Chattisgargh},
            country={India}}
             \vspace{-3mm}
\affiliation{organization={$^{4}$UNC Charlotte,},
            addressline={9201 University City Blvd}, 
            city={Charlotte},
            postcode={28223}, 
            state={NC},
            country={USA}}
             \vspace{-3mm}

\cortext[1]{Corresponding author}
\begin{abstract}
Optimal power flow (OPF) analysis in power system networks is challenging due to nonlinear power-flow relations and network characteristics. In addition, the high integration of distributed energy resources (DERs) adds an additional layer of complexity. Conventional nonlinear programming (NLP)-based OPF models suffer from infeasibility and computational challenges, with no guarantee of optimality. To address these issues, this article proposes an NLP-OPF model based on sequential quadratic programming to ensure an accurate solution for AC-DC distribution networks with high DER integration. The solution of the proposed model is compared with those of a convex SOCP-OPF model and an interior-point method OPF model to demonstrate its accuracy and performance. The proposed model is simulated and evaluated across multiple distribution systems, considering high DER integration scenarios. The analysis demonstrates that the proposed NLP-OPF model is scalable to large, extensive networks with a reasonable time to convergence (ToC), ensuring an accurate solution for AC-DC distribution networks. 
\end{abstract}

\begin{graphicalabstract}
\end{graphicalabstract}

\begin{highlights}
\item 
Beyond setting DERs' operational set-points, the proposed OPF method optimizes the modulation index of converters (i.e., AC/DC rectifiers and DC/AC inverters) using multi-objective functions to enhance both solution accuracy and reliability.

\item 
 To enhance the voltage stability margin (VSM), the objective functions incorporate a minimum voltage deviation condition weighted by a variable factor. Under the maximum VSM and minimum operational loss conditions, the system remains within operational limits over a wide range of loads supplied by the DERs. 
\item
The proposed NLP-SQP method is very fast and scalable with high integration of DERs.
\end{highlights}


\end{frontmatter}



\section{Introduction}
\label{sec:introduction}
In conventional power systems, electricity was predominantly generated using AC synchronous generators, while power transformers played a crucial role in efficiently transmitting electricity across various voltage levels. However, advancements in power electronics, the rise of distributed energy resources (DERs) \citet{6331779}, and the increasing integration of DC loads—such as digital devices, electric vehicles, and home appliances \citet{GERBER2019274}—are shifting the paradigm toward DC systems. Studies highlight DC’s advantages, including lower transmission losses, improved sustainability, and seamless integration of renewable energy \citet{9697309}. Additionally, government policies are accelerating the adoption of renewables \citet{TOMIN2022903}. However, a complete transition to DC is impractical, driving research toward hybrid AC-DC systems to harness the strengths of both \citet{9435084}. \par

Optimal Power Flow (OPF) is a valuable tool for distribution system planning to ensure the reliable, economical, and secure operation of power systems. It assists in decision-making for generation dispatch, resource scheduling, network planning, and market participation, leading to efficient utilization of resources and optimal system performance \citet{Intro3}. AC-OPF, which stands for Alternating Current Optimal Power Flow, is the commonly used term for OPF analysis  \citet{7480375}. However, due to the nonlinearity of the AC-OPF formulations, finding a feasible and accurate solution is not always guaranteed  \citet{7063278}. Linear approximations such as DC-OPF, based on simplified power flow equations, are often employed to address this issue, which compromises the accuracy of the solution \citet{linear3}. In recent times, convex OPF models (i.e., SOCP-OPF, SDP-OPF \citet{EES-012})  have gained interest due to their ability to provide exact solutions for radial distribution networks under some mild conditions. Nonetheless, these models still rely on relaxations and approximations that may reduce solution accuracy under certain conditions. \par 

\subsection{Existing OPF Analysis Methods for AC-DC Distribution Network}
Several OPF approaches exist for purely AC networks, but limited research has explored AC-DC distribution systems \citet{altorok2024review}. Ref. \citet{8636236} developed an AC-DC OPF model by transforming nonlinear AC-OPF formulations into convex and linearized models, including a convex relaxation of a parameterized AC-DC converter mode. A semidefinite relaxation of AC-OPF was proposed in \citet{7454786}, while \citet{10532381} optimized the hybrid AC-DC network topology based on critical node identification. The AC-DC OPF toolbox '\textit{pyflow}' for MTDC networks, introduced in \citet{10857445}, focuses solely on minimizing generation costs. However, these methods are designed for transmission systems and overlook DC loads and the network's radiality. For distribution systems, an ADMM-based optimization approach in \citet{10532334} minimizes real power loss and DER dispatch but does not address voltage deviations caused by converters. Additionally, the ADMM method converges slowly, often requiring over $\ge 100$ macro-iterations for small networks. Ref. \citet{10533268} presents a collaborative optimization method integrating converters and electric tourist buses to enhance network flexibility, primarily considering real power loss while neglecting general DC loads and BESS. \par

\subsection{Major Contributions}
From the above discussions, it is evident that there is a research gap in AC-DC OPF methods in distribution systems that i) consider high integration of DERs, ii) take a multi-objective optimization problem considering loss, generation cost, and voltage deviation (VD), iii) include DC loads, iv) are scalable and fast to converge. Again, due to limited nonlinear modeling capability, limited flexibility in constraint formulation, and computational complexity, the SOCP and SDP-based approach will not be suitable for efficient OPF analysis of the AC-DC distribution network (as this type of network has highly nonlinear properties due to the inclusion of converters and other control requirements). A comparison of recent OPF methodologies for the AC-DC network is provided in Table \ref{research_gap}. \par
\begin{table}[]
\centering
\caption{\textsc{A Comparison of Recent OPF Methodologies for the AC-DC Network (NR: Not Reported)}}
\label{research_gap}
\begin{tabular}{ccccccc}
\hline\hline
Ref. & Radiality & DERs & BESS & \begin{tabular}[c]{@{}c@{}}DC \\ loads \end{tabular} & VD &ToC \\ \hline
  \citet{8636236}   &  $\times$         &   $\times$    &   $\times$    &    $\times$  &  $\times$   &  fast   \\ \hline
  \citet{7454786}   &   $\times$         &   \checkmark     &   $\times$   &     $\times$  &  $\times$    &  NR   \\ \hline
  \citet{10532381} &    $\times$        &  $\times$     &   $\times$    &     $\times$  &  $\times$   &   NR  \\ 
  \hline
   \citet{10857445}  &   $\times$         &   \checkmark    &   $\times$    &  $\times$    &  $\times$    &  NR   \\ \hline
   \citet{10532334}  &    \checkmark        &   \checkmark     &   \checkmark     &  \checkmark    &   $\times$    &   very slow  \\ \hline
   \citet{10533268}  &     \checkmark       &  $\times$     &   $\times$   &  \checkmark    &  $\times$    &  NR   \\ \hline
\begin{tabular}[c]{@{}c@{}}Proposed \\ method \end{tabular}    &  \checkmark         &   \checkmark    &   \checkmark    &     \checkmark  &  \checkmark  &   very fast  \\ \hline\hline
\end{tabular}
\end{table}
From this point, for the AC-DC grids with converters and DERs, nonlinear programming (NLP)-based methods, especially the Sequential Quadratic Programming (NLP-SQP) algorithm, are more reliable and robust frameworks for OPF analysis \citet{Tolle_sqp}. The reasons include, but are not limited to, modeling flexibility, solution accuracy, the incorporation of nonlinear effects, and better addressing of system complexity. The efficiency and convergence criteria of the NLP-SQP algorithm with significantly constrained optimization are discussed in \citet{Tolle_sqp}. However, this convergence criterion needs to be extended for the power distribution network. In this paper, we propose a multi-objective OPF methodology for AC-DC distribution networks that addresses the research gaps mentioned above. The proposed AC-DC OPF models consider the active/reactive powers and voltage magnitudes of converters at PCCs and AC/DC buses as independent control variables. The main contributions of this approach include: \par
 
\begin{itemize}
\item  Beyond setting DERs' operational set-points, the proposed OPF method optimizes the modulation index of converters (i.e., AC/DC rectifiers and DC/AC inverters) using multi-objective functions to enhance both solution accuracy and reliability.
 \item To enhance the voltage stability margin (VSM), the objective functions incorporate a minimum voltage deviation condition weighted by a variable factor. Under the maximum VSM and minimum operational loss conditions, the system remains within operation limits over a wide range of loads supplied by the DERs. 
\item The proposed NLP-SQP method is very fast and scalable with high integration of DERs.
 \end{itemize}
The structure of this paper is as follows: Section~\ref{AC-DC network model} presents the AC-DC distribution network, while Section~\ref{OPF-model} outlines the proposed OPF methodology. Section~\ref{solution-opf} and Section~\ref{convergence-criteria} discuss the solution methodology and the convergence criteria of the OPF problem, respectively. Section~\ref{simulation-result} presents the analysis of results from the test case simulation, followed by the conclusion in Section~\ref{conclusion}.

\section{AC-DC Distribution Network} \label{AC-DC network model}
\begin{figure}[h!]
        \centering
        \includegraphics[trim=0mm 0mm 0mm 0mm,,width=8cm,height=5.0cm]{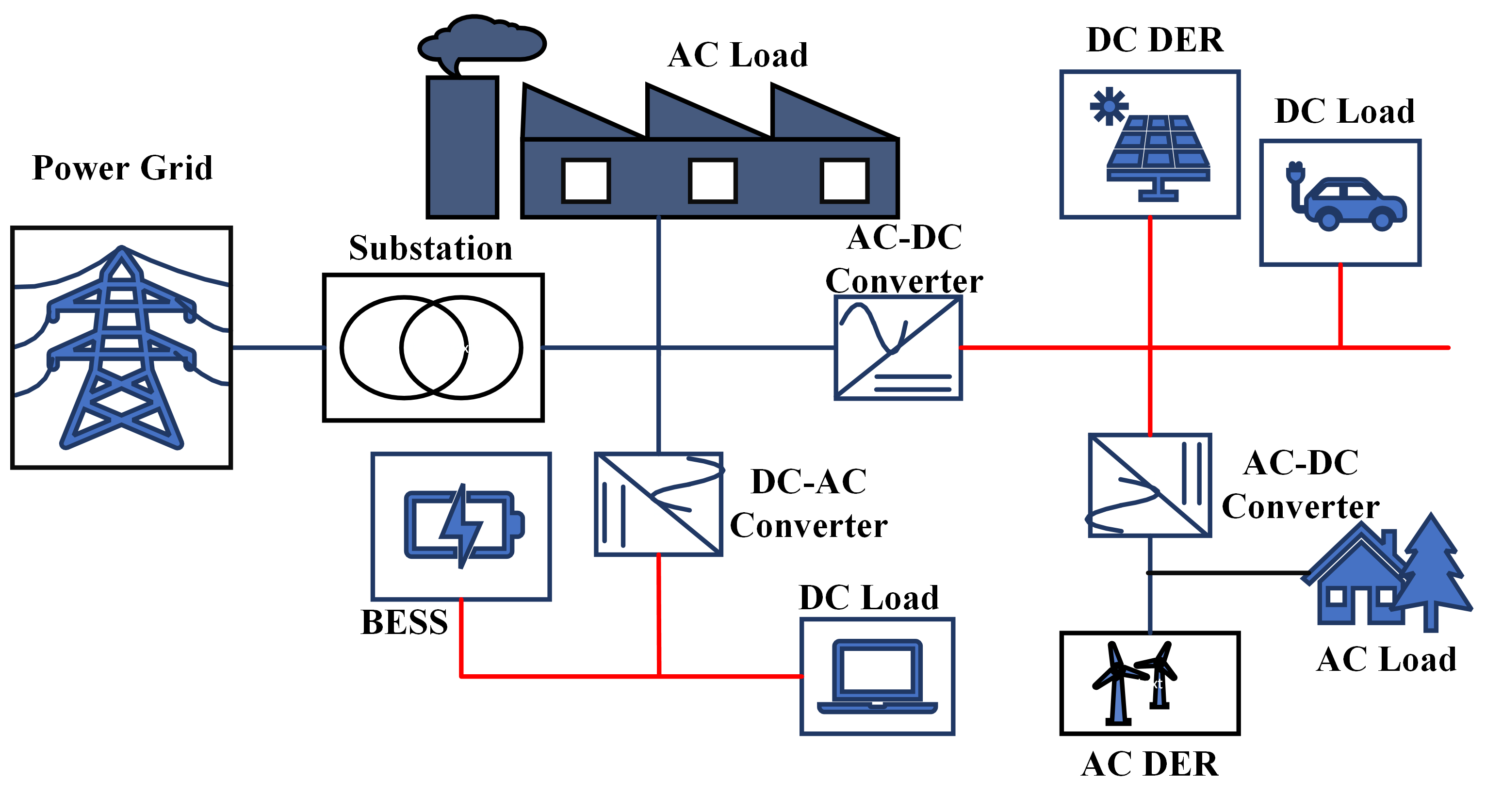}
        \caption{A modern AC-DC distribution system.} 
        \label{fig1}   
    \end{figure}

AC-DC distribution systems differ structurally from AC distribution systems as they integrate AC and DC components, including loads, generating units, lines, and buses. In the OPF formulation, we have incorporated the mathematical models of DC distribution lines,  DC loads, DC buses, and various types of DC distributed generations with standard AC components. Fig. \ref{fig1} represents a simple AC-DC distribution system diagram.  Ref. \citet{pu1} comprehensively detailed the mathematical models for power converters, loads, AC, and DC lines. Accurate mathematical modeling of power converters is crucial for OPF analysis. This study adopts the per-unit equivalent model of power converters as presented in \citet{pu1}.\par 
     
\noindent The input and output terminal of the  converters' per-unit voltage and power relationship can be represented, respectively, as:   
\vspace{-4mm}
\begin{equation}
    \label{eq1}
     V_{j}=MV_{i}
\end{equation}
\vspace{-8mm}
\begin{equation}
    \label{eq2}
     P_{out}= \eta P_{in} 
\end{equation}
where $V_i$ and $V_j$ are the converters' input and output terminal voltage, respectively. Moreover, $P_{in}$ and $P_{out}$ are the converters' input and output real power, respectively.  $\eta $ is the efficiency of the respective converters, and  $M$ is the modulation index. The minimum and maximum modulation index of the converter is $M^{min} = \frac{V^{min}}{V^{max}}$ and $M^{max} = \frac{V^{max}}{V^{min}}$.
The converters' input and output side base value relationship for voltage $V_{dc} = \frac{2\sqrt{2}}{\sqrt{3}}V_{ac}$ and $V_{ac} = \frac{\sqrt{3}}{2\sqrt{2}}V_{dc}$ for AC-DC and DC-AC converter, respectively.

\section{Proposed Optimal Power Flow Model}\label{OPF-model}
\subsection{Modeling of Power System Network}
Considering a distribution system with constant complex loads $s^d_i = p^d_i +jq^d_i$ connected to bus $i\in \mathcal{N}$ where $p^d_i$ and $q^d_i$ are real and reactive power load and $\mathcal{N}$ is the set of all buses, respectively. Moreover, $G_{ii}$ and $B_{ii}$ are the conductance and susceptance connected to the bus $i \in \mathcal{N}$, dependent on voltage magnitude. Line impedance of a branch $L_{ij} \in \mathbf{L}$ is designed as $z_{ij} = r_{ij} + jx_{ij}; j=\sqrt{-1}$. Here, $\mathbf{L}$ is the set of all branches in the network. 
The active and reactive power balance equations at bus $i \in \mathcal{N}$ is as follows: 
\begin{subequations}
\label{eqn2}
\begin{equation}
\label{eqn2a}
   p^g_i - p^d_i -V_i^2G_{ii} -  \sum_{j=1, j\ne i}^nV_iV_j[G_{ij}cos(\theta_i-\theta_j)\nonumber\\ 
   + B_{ij}sin(\theta_i-\theta_j)] = 0
\end{equation}
\vspace{-4mm}
\begin{equation}
    \label{eqn2b}
    q^g_i - q^d_i +V_i^2B_{ii} - \sum_{j=1,j\ne i}^nV_iV_j[G_{ij}sin(\theta_i-\theta_j)\nonumber\\
    - B_{ij}cos(\theta_i-\theta_j)] = 0
\end{equation}
\end{subequations}

\noindent Here, $V_i$ and $V_j$ denote the voltage of bus $i, j \in \mathcal{N}$, where $p^g_i$ and $q^g_i$ are the active and reactive power injection by DERs to the corresponding bus $i \in \mathcal{N}$. Moreover, $G_{ij} = \frac{r_{ij}}{r_{ij}^2+x_{ij}^2}$, and $B_{ij} = \frac{x_{ij}}{r_{ij}^2+x_{ij}^2}$, are conductance and susceptance,  respectively. \par

If there is a converter at bus $j$, we are creating a dummy bus at the converter's input side as $j^{'}$. The converter's input and output relationship can be formulated as follows: 

\begin{equation}
    \label{eqn03}
    V_j = V_{j'} M
\end{equation}

\subsection {Objectives of the optimal power flow model} 
The main goal of optimal power flow is to find the optimal operating point, that is, $p_i^g, q_i^g$, etcetera, to balance the supply and demand. For this work, based on a required objective function $f$ and maintaining the operational and control constraints, we have considered different objectives such as loss minimization, generator cost minimization, and voltage deviation minimization. The objective functions for the AC-DC network are established as follows: 
\begin{equation}
    \label{eqn12}
    f = min \sum f_{AC}(x^{AC},y^{AC}) + f_{DC}(x^{DC},y^{DC}) 
\end{equation}

\noindent Here, $x$ represents the vector of controllable variables, including active and reactive power from controllable shunts, while $y$ denotes the vector of state variables, such as voltage magnitude. The objective function $f$ comprises two components: $f_{AC}$ and $f_{DC}$, corresponding to the AC and DC parts.

(a) real power generation cost minimization,
\begin{equation}
    \label{eqn13}
     f_{1}= min \sum_{i\in N_{g}} [c_0^i + c_1^ip_i^g + c_2^i(p_i^g)^2]
\end{equation}

or Network real power loss minimization,
\begin{equation}
    \label{eqn14}
  f_1 = min \sum_{i}\sum_{j}|P_{ij} - P_{ji}|=min \sum_{i}\sum_{j}P_{ij}^{L}
\end{equation}
\noindent where $c_0^i$, $c_1^i$, and $c_2^i$ represent the generator's quadratic cost coefficients at bus $i\in N_g$. $P_{ij}$ denotes the real power from bus $i$ to bus $j$, while $P_{ji}$ represent the injected real power at bus $j$ from bus $i$. $P_{ij}^{L}$ is the real power loss of line $L_{ij}\in \textbf{L}$. \par
b) Our proposed article considers the converters for the OPF model, which supports the reactive power. Consequently, converters tend to operate at their maximum reactive power support limits to minimize costs or network losses, resulting in their maximum voltage limits. This implies that the power system may lack sufficient reserves to supply reactive power during contingencies under specific operating conditions \citet{article33}. One approach to addressing this issue is incorporating voltage deviation into the objective function \citet{article34}. The voltage deviation from its reference value can be formulated as an objective function as follows: 
\begin{equation}
    \label{eqn15}
    {f_2} = min \sum_{i\epsilon N} {\frac{|V_i^2 - ({V_i}^{ref})^2|}{N}}   
\end{equation}
\noindent where ${V_i}^{ref}$ represents the voltage reference for bus $i \in \mathcal{N}$. As such, the objective is considered as the combination of $f_1$ and $f_2$, which is defined as:
\begin{align}
\label{eqn16}
{f} = {(1-K)f_1} + {Kf_2}; \; \;  K\: \mathlarger{\epsilon} \; [0,1]
\end{align}
Where $K$ is the weight factor for each objective, based on the value of $K$, the weight of each objective changes in \eqref{eqn16}, and the voltage deviation varies. \par

\subsection{OPF model with Network Operational Constraints}
For the AC-DC network OPF analysis, the converter modulation index $(M)$ is also considered an operational variable. For the objectives (\ref{eqn16}) discussed in this article, besides the power flow equality constraints (\ref{eqn2a}),   the following inequality constraints are considered for the network analysis. $\underline{(.)}$ and $\overline{(.)}$ denote the minimum and maximum bounds of the variables. 
    \begin{align}
        \label{eqn3a}
        \begin{cases}
    \underline{\theta}_{i}  \le \theta_i \le  \overline{\theta}_{i}\\ 
        \underline{V}_{i} \le V_i \le  \overline{V}_{i}\\ 
        \underline{p}^g_{i} \le p^g_i \le  \overline{p}^g_{i}\\
    \underline{q}^g_{i}\ \le q^g_i \le  \overline{q}^g_{i}\\ 
        \underline{M_{i}} \le M_{i} \le  \overline{M_{i}}
        \end{cases}
    \end{align}
\begin{figure}[h!]
    \centering
    \includegraphics[trim=0mm 0mm 0mm 0mm,,width=8.5 cm,height=8.6cm]{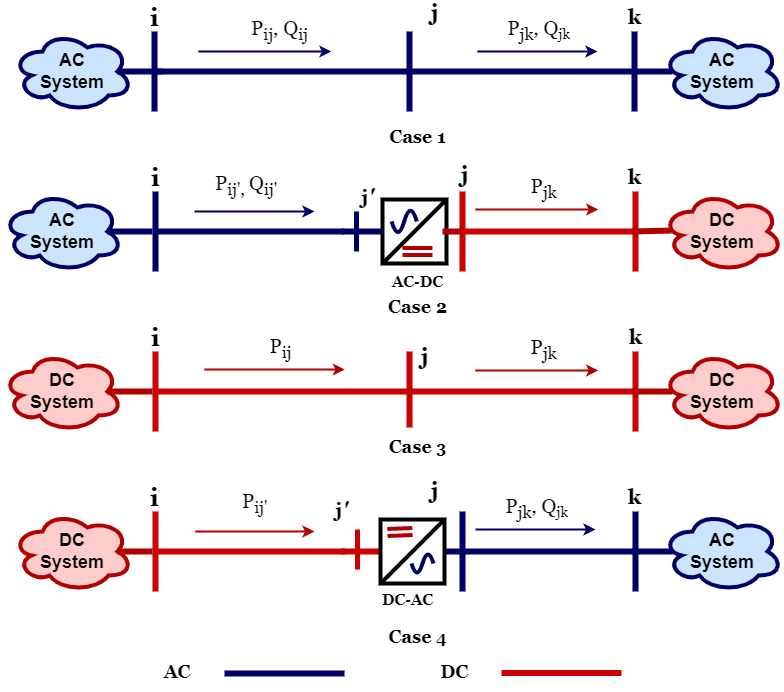}
    \caption{AC-DC network model with converters.}
    \vspace{-5.5mm}
    \label{Figure1}

\end{figure}

\subsection {OPF Model of AC-DC Distribution Network} 
An AC-DC distribution network comprises both an AC network and a DC network. An AC-DC power distribution network can have the following four line configurations shown in Fig. \ref{Figure1}. This section will formulate the OPF equations for the AC-DC network for the four cases illustrated in Fig.~\ref{Figure1}.\par
\subsubsection{ \textit{Case 1 (AC - AC Network)}} \label{AC-AC}
For this case, only the AC sections of the AC-DC network are taken into account. (\ref{eqn16}) will be the objective function. However, as the system is an AC network, there will not be a converter input-output voltage conversion relation, (\ref{eqn03}) will not be in the equality constraints, and the modulation index ($M$) limit will not be in inequality constraints.  The unknown variables of this problem are $(\theta_i, V_i,p^g_i,q^g_i)$. So the OPF's constraint equations will be (\ref{eqn2}) and  (\ref{eqn3a}) without the modulation index ($M$) limit. 
\par

\subsubsection{\textit{Case 2 (AC - DC Converter)}} \label{AC-DC}
This case is analyzed at the coupling buses where an AC section connects to a DC section via an AC-DC converter. An example of such a converter is illustrated in Fig. \ref{Figure1}. The AC-DC converter is placed at bus $j$, with a corresponding dummy bus $j'$ created at its input side. Defining the converter’s modulation index as $M_{AC-DC}$, the power flow through branch ${L}_{ij'}$ before the converter can be expressed as:
\begin{subequations}
    \label{eqn5}
    \begin{equation}
        \label{eqn5a}
        P^{AC}_{in}  = \sum_{i:i\xrightarrow{}j} (P_{ij'} - r_{ij'}|I_{ij'}|^2)
    \end{equation}
    \vspace{-5mm}
    \begin{equation}
        \label{eqn5b}
        Q^{AC}_{in}  = \sum_{i:i\xrightarrow{}j} (Q_{ij'} - x_{ij'}|I_{ij'}|^2)
    \end{equation}
\end{subequations}
Where $P^{AC}_{in}$ is the input power and $P^{DC}_{out}$ output power of the converter. So, the voltage and power relationship for the proposed model will be: 
\begin{subequations}
    \label{eqn6}
    \begin{equation}
        \label{eqn6a}
        P^{DC}_{out} = \eta P^{AC}_{in}
    \end{equation}
    \vspace{-5mm}
    \begin{equation}
        \label{eqn6b}
        V_{j} = M_{AC-DC}V_{j'}
    \end{equation}
\end{subequations}

\par
So, (\ref{eqn16}) will be the objective function, and the constraints equation will be (\ref{eqn3a}) and (\ref{eqn6}).
\par
\subsubsection{\textit{Case 3 (DC-DC Network)}} \label{DC-DC}
Only the DC sections of the AC-DC distribution network are taken into account. So, the real power part of (\ref{eqn2}) will be considered for equality constraint for this case. Our objective function does not consider reactive power loss or the cost of reactive power, so the objective function will not change from case $1$ and $2$. The unknown variables will be changed to $ [V_i, p^g_i]$.  So the inequality constraints equation for this case will be the voltage angle, voltage magnitude, and real power limits of (\ref{eqn3a}). 
\par
\subsubsection{ \textit{Case 4 (DC-AC Converter)}} \label{DC-AC}
This applies to coupling buses where a DC section connects to an AC section via a DC-AC converter. A DC-AC network is illustrated in case 4 of Fig. \ref{Figure1}. A dummy bus is created on the input side of the DC-AC converter, as in Case 2. The dummy bus is denoted as $j'$. So, the active power flow through the line ${L}_{ij'}$, that is just before the converter will be: 
\begin{equation}
    \label{eqn9}
    P^{DC}_{in}  = \sum_{i:i\xrightarrow{}j} (P_{ij'} - r_{ij'}|I_{ij'}|^2)
\end{equation}
Where $P^{DC}_{in}$ is the real power input to the converter. If $M_{DC-AC}$ is the modulation index of the converter, the power flowing out of the converter and the voltage of the bus $j$ will be: 
\begin{subequations}
    \label{eqn10}
    \vspace{-4mm}
    \begin{equation}
        \label{eqn10a}
        P^{AC}_{out} = \eta P^{DC}_{in}
    \end{equation}
    \vspace{-8mm}
    \begin{equation}
        \label{eqn10b}
        V_{j} = M_{DC-AC}V_{j'}
    \end{equation}
\end{subequations}
Where $P^{AC}_{in}$ is the output active power to the converter.
The objective function of the OPF problem will be as before. However, equations (\ref{eqn3a}) and (\ref{eqn10}) will form the constraints.

\section{Solution of the OPF Problem} \label{solution-opf}
For solving the OPF problems from (\ref{eqn2}) to (\ref{eqn3a}) with SQP, the OPF formulations can be written as follows, considering real power cost minimization. For writing simplicity, we will consider the equality constraint equations as $h(.)$ and inequality constraints as $g(.)$ as this will change based on case 1 to case 4. 
\begin{subequations}
    \label{equation9}
    \begin{equation}
    \begin{split}
        \label{equation9a}
        \min(1-K)\sum_{i\in N_{g}} [c_0^i + c_1^ip_i^g + c_2^i(p_i^g)^2] + K\sum_{i\in N} {\frac{|V_i^2 - ({V_i}^{ref})^2|}{N}} 
    \end{split}
    \end{equation}
    \vspace{-5mm}
    \begin{equation}
      \label{equation9d}
      s.t.,
      \begin{cases}
        h(\begin{matrix}[\theta_{i}& V_{i}& M_{i}]^{T}          
        \end{matrix}) = 0\\ 
        g(\begin{matrix}[\theta_{i}& V_{i}& p_i^g& q_i^g& M_{i}]^{T}         \end{matrix}) \le 0
       \end{cases}
    \end{equation}
\end{subequations}
Where (\ref{equation9a}) and (\ref{equation9d}) represent the objective function and constraints of the OPF problem, respectively. The OPF problem in (\ref{equation9}) can be replaced with its quadratic approximations in the following way: 
\begin{subequations}
    \label{equation10}
    \begin{equation}
    \begin{split}
     \label{equation10a}
     \min\begin{bmatrix}
         (1-K)\sum_{i\in N_{g}}[c_{i}^{i}+2c_{2}^{i}(p_{i}^{g})^{m}]\\ K\sum_{i\in N}\frac{2(V_{i})^{m}}{N}
     \end{bmatrix} \begin{bmatrix}
         V_{i}-(V_{i})^{m}\\ p_{i}^{g}-(p_{i}^{g})^{m}
     \end{bmatrix}
     +\frac{1}{2} \begin{bmatrix}
         V_{i}-(V_{i})^{m}\\ p_{i}^{g}-(p_{i}^{g})^{m}
     \end{bmatrix}^{T} \\
     \begin{bmatrix}
         (1-K)\sum_{i\in N_{g}}2c_{2}^{i} & 0 \\ 0 & K\sum_{i\in N}\frac{2}{N}
     \end{bmatrix}  \begin{bmatrix}
         V_{i}-(V_{i})^{m}\\ p_{i}^{g}-(p_{i}^{g})^{m}
     \end{bmatrix}
    \end{split}
    \end{equation}
    \begin{equation}
    \label{equation10d}
    s.t.,
    \begin{cases}
     h(\begin{bmatrix}\theta_{i}\\ V_{i}\\ M_{i}          
        \end{bmatrix}) + \nabla h(\begin{bmatrix}\theta_{i}\\ V_{i}\\ M_{i}\end{bmatrix}) (\begin{bmatrix}\theta_{i}-\theta_{i}^{m}\\ V_{i}-V_{i}^{m}\\ M_{i} -M_{i}^{m}\end{bmatrix}=0
        \\ 
        g(\begin{bmatrix}\theta_{i}\\ V_{i}\\ p_i^g\\ q_i^g\\ M_{i}\end{bmatrix}) + \nabla g(\begin{bmatrix}\theta_{i} \\ V_{i} \\ p_i^g \\ q_i^g\\ M_{i}\end{bmatrix}) (\begin{bmatrix}\theta_{i}-\theta_{i}^{m}\\V_{i}- V_{i}^{m}\\ p_i^g-(p_i^g)^{m}\\ q_i^g-(q_i^g)^{m}\\ M_{i}- M_{i}^{m} 
        \end{bmatrix}  + \delta_{z} = 0  
    \end{cases}
    \end{equation}
    where,
    \begin{equation}
    \begin{split}
        \delta_{z}=-\nabla g(\begin{bmatrix}
           \theta_{i}^{m}\\ V_{i}^{m}\\ (p_i^g)^{m}\\ (q_i^g)^{m}\\ M_{i}^{m} \end{bmatrix}) (\begin{bmatrix}
           \theta_{i}-\theta_{i}^{m}\\ V_{i}-V_{i}^{m}\\ p_i^g-(p_i^g)^{m}\\q_i^g-(q_i^g)^{m}\\ M_{i}-M_{i}^{m} 
        \end{bmatrix} - g(\begin{bmatrix}\theta_{i}^{m}\\ V_{i}^{m}\\ (p_i^g)^{m}\\ (q_i^g)^{m}\\ M_{i}^{m}  \end{bmatrix}) -z^{m}     
    \end{split}
    \end{equation}
\end{subequations}

\noindent Where $\nabla f(x)$ and $Hf(x)$ represent the gradient and Hessian of $f(x)$ at $x \in \textit{R}^n$. A slack variable $z, (z\in \textit{R}^n)$ is added in the inequality constraints (\ref{equation10d}). The vector $[\theta_{i}\; V_{i}\; (p_i^g)\; (q_i^g)\; M_{i}]^{T} -[\theta_{i}^{m}\; V_{i}^{m}\; (p_i^g)^{m}\;(q_i^g)^{m}\; M_{i}^{m}]^{T}$ and $z-z^{m}$ in the above equations give the step direction for $m^{th}$ iteration. The update for each iteration will be: 
\begin{subequations}
    \label{equation11}
    \begin{equation}
        \label{equation11a}
        \begin{bmatrix}
          \theta_{i}^{m+1}\\ V_{i}^{m+1}\\ (p_i^g)^{m+1}\\(q_i^g)^{m+1}\\ M_{i}^{m+1}  
        \end{bmatrix}= \begin{bmatrix} \theta_{i}^{m}\\ V_{i}^{m}\\ (p_i^g)^{m}\\(q_i^g)^{m}\\ M_{i}^{m}\end{bmatrix}+ \alpha \begin{bmatrix}
         \theta_{i}-\theta_{i}^{m}\\ V_{i}-V_{i}^{m}\\ p_i^g-(p_i^g)^{m}\\q_i^g-(q_i^g)^{m}\\ M_{i}-M_{i}^{m}
        \end{bmatrix} 
    \end{equation}
    \begin{equation}
        \label{equation11a}
        z^{m+1} = z^m + \alpha (z-z^m )
    \end{equation}
\end{subequations}
$\alpha$ is the step direction parameter whose values is typically between $(0,1]$ \citet{Tolle_sqp}. The iteration stops when the following criteria are satisfied, and the solution is optimal. 
\begin{subequations}
    \begin{equation}
    \begin{split}
        |g(\begin{bmatrix}\theta_{i}^{m}\\ V_{i}^{m}\\ (p_i^g)^{m}\\ (q_i^g)^{m}\\ M_{i}^{m}]  \end{bmatrix})|, \:
        \frac{\begin{bmatrix}\theta_{i}^{m}\\ V_{i}^{m}\\ (p_i^g)^{m}\\ (q_i^g)^{m}\\ M_{i}^{m}  \end{bmatrix}- \begin{bmatrix}\theta_{i}^{m-1}\\ V_{i}^{m-1}\\ (p_i^g)^{m-1}\\ (q_i^g)^{m-1}\\ M_{i}^{m-1}]  \end{bmatrix}}{1+|\begin{bmatrix}\theta_{i}^{m}\\ V_{i}^{m}\\ (p_i^g)^{m}\\ (q_i^g)^{m}\\ M_{i}^{m}]  \end{bmatrix}|} < \epsilon
    \end{split}
    \end{equation}
\end{subequations}
\noindent here, $\epsilon$ is a predefined tolerance limit. The formulation for the real power loss minimization can be formed in a similar way by replacing the first part of \eqref{equation9a} with \eqref{eqn14}.

\section{Convergence Analysis} \label{convergence-criteria}

For convergence of the proposed method,
for a constant $\mathcal{K}>0$,   $([\theta_{i}^{m} \quad V_{i}^{m}$$ \quad (p_i^g)^{m} $  $\quad (q_i^g)^{m}\quad M_{i}^{m} ]^{T},z^m) \in \mathcal{G}$, there exist a solution of (\ref{equation10}) such that $||[\theta_{i}\; V_{i}\; (p_i^g)\; (q_i^g)\; M_{i}]^{T} -[\theta_{i}^{m}\; V_{i}^{m}\; (p_i^g)^{m}\;(q_i^g)^{m}\; M_{i}^{m}]^{T}, \mu^m|| \le \mathcal{K}$, where $\mathcal{G}$ is a compact set of $\textit{R}^n\times\textit{R}_+^n$ and $\mu^m$ is a multiplication factor. Moreover, $\mathcal{S}$ is the finite first-order solution set, and $C_{\eta}$ is the $\eta$th neighborhood of the feasible set where $(C, \mathcal{G}) \subset \mathcal{S}$.\par


For convergence, it is needed to prove that any iterate $\{([\theta_{i}^{m}\quad V_{i}^{m}\quad (p_i^g)^{m}\quad $ $(q_i^g)^{m}\quad M_{i}^{m}]^{T},z^m)\}$ converges to a point $(\begin{bmatrix}\theta_{i}^{*}& V_{i}^{*}& (p_i^g)^{*}& (q_i^g)^{*}& M_{i}^{*} \end{bmatrix}^{T},z^*) \in \mathcal{S}$ where $\begin{bmatrix}\theta_{i}^{*}& V_{i}^{*}& (p_i^g)^{*}& (q_i^g)^{*}& M_{i}^{*} \end{bmatrix}^{T}$ is the first order solution of (\ref{equation10}). 
First, it is needed to prove $([\theta_{i}\; V_{i}\; (p_i^g)\; (q_i^g)\; M_{i}]^{T} -[\theta_{i}^{m}\; V_{i}^{m}\; (p_i^g)^{m}\;(q_i^g)^{m}\; M_{i}^{m}]^{T}, z-z^m) \to 0$. For this, considering $\eta^*\ge 0$ is the smallest value of $\eta;  \forall \: C_{\eta^*}$, and the following condition must be satisfied: 
\begin{subequations}
    \label{equation20}
    \begin{equation}
    \begin{split}
        \label{equation20a}
        \phi(\begin{bmatrix}\theta_{i}^{m}\\ V_{i}^{m}\\ (p_i^g)^{m}\\ (q_i^g)^{m}\\ M_{i}^{m} \end{bmatrix}+\alpha \begin{bmatrix}
           \theta_{i}-\theta_{i}^{m}\\ V_{i}-V_{i}^{m}\\ p_i^g-(p_i^g)^{m}\\q_i^g-(q_i^g)^{m}\\ M_{i}-M_{i}^{m} 
        \end{bmatrix}, z^m + \alpha z-z^m) \leq \phi(\begin{bmatrix}\theta_{i}^{m}\\ V_{i}^{m}\\ (p_i^g)^{m}\\ (q_i^g)^{m}\\ M_{i}^{m} \end{bmatrix}, z^m) \\+ \frac{1}{2}\alpha \sigma \nabla \phi(\begin{bmatrix}\theta_{i}^{m}\\ V_{i}^{m}\\ (p_i^g)^{m}\\ (q_i^g)^{m}\\ M_{i}^{m} \end{bmatrix}, z^m)^T(\begin{bmatrix}
           \theta_{i}-\theta_{i}^{m}\\ V_{i}-V_{i}^{m}\\ p_i^g-(p_i^g)^{m}\\q_i^g-(q_i^g)^{m}\\ M_{i}-M_{i}^{m} 
        \end{bmatrix}, z-z^m)
        \end{split}
    \end{equation}
    \vspace{-4mm}
    \begin{equation}
        \label{equation20b}
        \nabla\phi(\begin{bmatrix}\theta_{i}^{m}\\ V_{i}^{m}\\ (p_i^g)^{m}\\ (q_i^g)^{m}\\ M_{i}^{m} \end{bmatrix},z^m)^T(\begin{bmatrix}
           \theta_{i}-\theta_{i}^{m}\\ V_{i}-V_{i}^{m}\\ p_i^g-(p_i^g)^{m}\\q_i^g-(q_i^g)^{m}\\ M_{i}-M_{i}^{m} 
        \end{bmatrix}, z-z^m) \leq -v||\begin{bmatrix}
           \theta_{i}-\theta_{i}^{m}\\ V_{i}-V_{i}^{m}\\ p_i^g-(p_i^g)^{m}\\q_i^g-(q_i^g)^{m}\\ M_{i}-M_{i}^{m} 
        \end{bmatrix}, z-z^k||^2
    \end{equation}
\end{subequations}
The value of $\alpha$ and $\sigma$ is $(0,1]$ and $(0, 1/2)$, respectively. If $v$ is very small, such that, $v \to 0$, it can be written $\phi(\begin{bmatrix}\theta_{i}^{m+1}& V_{i}^{m+1}& (p_i^g)^{m+1}& (q_i^g)^{m+1}& M_{i}^{m+1} \end{bmatrix}^{T}) \leq \phi(\begin{bmatrix}\theta_{i}^{m}& V_{i}^{m}& (p_i^g)^{m}& (q_i^g)^{m}& M_{i}^{m} \end{bmatrix}^{T})$, which implies $\{[\theta_{i}\; V_{i}\; (p_i^g)\; (q_i^g)\; M_{i}]^{T} -[\theta_{i}^{m}\; V_{i}^{m}\; (p_i^g)^{m}\;(q_i^g)^{m}\; M_{i}^{m}]^{T}, z-z^m\} \to 0$. These criteria have to be fulfilled to get a tractable solution.

\section{Simulation and Result Analysis} \label{simulation-result}
The proposed AC-DC method has been implemented using MATLAB's Optimization Toolbox on a system equipped with an Intel(R) Core(TM) i7-3770 CPU @ 3.40 GHz processor and 16GB RAM. This OPF model determines the optimal operating points $(p^g_i, q^g_i)$ and the optimal modulation index $M$ for AC-DC radial distribution networks. It is tested on two test cases: a modified 32-bus distribution network (Test Case I) and the IEEE 123-bus distribution network (Test Case II).
To validate the proposed AC-DC network, a power flow analysis is performed by setting the objective function (\ref{eqn16}) to zero and comparing the results with the graph-theoretic-based AC-DC power flow model from \citet{pu1}. For Test Case I, the maximum voltage profile error in the base case using the NLP-SQP method is 0.8\%. The results of the proposed method are further compared with the SOCP-OPF model from \citet{10856679} and the modified IPM-based method from \citet{7906617}. A detailed analysis of these observations is presented in the following subsections, and the overall algorithm is outlined in Algorithm~\ref{algorithm_graph}.
\begin{algorithm}
 {Load system data, determine $\mathcal{N}$, $\mathbf{L}$, initialize  OPF variables $\mathbf{X}$ and set $\epsilon = 10^{-5}$.} \\
  {\For{$k=1:\mathcal{N}$}
  {
        \If{$L_{ij}(k)\in \mathbf{L}$ is a fully AC branch}
        {Run NLP-SQP-OPF using (\ref{AC-AC})}
        \If{an AC-DC converter is placed at the branch $L_{ij}(k)\in \mathbf{L}$}
        {
        Place a dummy bus as $j^{'}$ at the input side of the converter\\
        Run NLP-SQP-OPF using (\ref{AC-DC})}
        \If{$L_{ij}(k)\in \mathbf{L}$ is a fully DC branch}
        {
        Run NLP-SQP-OPF using (\ref{DC-DC})}
        \If{an DC-AC converter is placed at the branch $L_{ij}(k)\in \mathbf{L}$}
        {
        Place a dummy bus as $j^{'}$ at the input side of the converter\\
        Run NLP-SQP-OPF using (\ref{DC-AC})}
        }
  }
$\mathbf{X}$ returns the NLP-SQP-OPF results for all variables.
\caption{AC-DC NLP-SQP-OPF Methodology}
\label{algorithm_graph}
\end{algorithm}
\begin{table}[htbp]
\centering
\caption{\textsc{Optimal 'M' and Percent of the Network Loss for the 32-bus Network}}
\label{tab:Table IV}

\resizebox{\linewidth}{!}{

\begin{tabular}{c|c|c|c|c|c|c|c|c|c|c|c|c}
\hline\hline
Network & \begin{tabular}[c]{@{}c@{}}$V_{base}$\\ (AC-KV)\end{tabular} & \begin{tabular}[c]{@{}c@{}}$V_{base}$\\ (DC-KV)\end{tabular} & \begin{tabular}[c]{@{}c@{}}Weight\\ Factor\\ 'K'\end{tabular} & \begin{tabular}[c]{@{}c@{}}Conv.\\ Con.\\ Nodes\end{tabular} & \begin{tabular}[c]{@{}c@{}}Conv. \\ Type\end{tabular} & \begin{tabular}[c]{@{}c@{}}Conv.\\ ID\end{tabular} & \begin{tabular}[c]{@{}c@{}}Optimal\\ 'M'\\ (SQP)\end{tabular} & \begin{tabular}[c]{@{}c@{}}Optimal\\ 'M'\\ (IPM)\end{tabular} & \begin{tabular}[c]{@{}c@{}}Optimal\\ 'M'\\ (SOCP)\end{tabular} & \begin{tabular}[c]{@{}c@{}}Network\\ Loss (\%)\\ (SQP)\end{tabular} & \begin{tabular}[c]{@{}c@{}}Network\\ Loss (\%)\\ (IPM)\end{tabular} & \begin{tabular}[c]{@{}c@{}}Network \\ Loss (\%)\\ (SOCP)\end{tabular} \\ \hline

\multirow{9}{*}{\begin{tabular}[c]{@{}c@{}}32-bus\\ (Base \\ Case)\end{tabular}} & \multirow{9}{*}{12.66} & \multirow{9}{*}{17.90} & \multirow{3}{*}{0.0} & 6-7 & AC-DC & $M_1$ & 1.7686 & 1.7683 & 1.7524 & \multirow{3}{*}{8.437\%} & \multirow{3}{*}{8.436\%} & \multirow{3}{*}{8.967\%} \\ \cline{5-10}
 & & & & 12-13 & DC-AC & $M_2$ & 0.6219 & 0.6218 & 0.6235 & & & \\ \cline{5-10}
 & & & & 26-27 & AC-DC & $M_3$ & 1.7717 & 1.7614 & 1.7556 & & & \\ \cline{4-13}
 & & & \multirow{3}{*}{0.5} & 6-7 & AC-DC & $M_1$ & 1.6746 & 1.6729 & 1.6678 & \multirow{3}{*}{8.628\%} & \multirow{3}{*}{8.628\%} & \multirow{3}{*}{9.115\%} \\ \cline{5-10}
 & & & & 12-13 & DC-AC & $M_2$ & 0.6190 & 0.6202 & 0.6206 & & & \\ \cline{5-10}
 & & & & 26-27 & AC-DC & $M_3$ & 1.6751 & 1.6760 & 1.6667 & & & \\ \cline{4-13}
 & & & \multirow{3}{*}{0.9} & 6-7 & AC-DC & $M_1$ & 1.6813 & 1.6814 & 1.6618 & \multirow{3}{*}{8.658\%} & \multirow{3}{*}{8.757\%} & \multirow{3}{*}{\begin{tabular}[c]{@{}c@{}}9.581\%\end{tabular}} \\ \cline{5-10}
 & & & & 12-13 & DC-AC & $M_2$ & 0.6195 & 0.6200 & 0.6208 & & & \\ \cline{5-10}
 & & & & 26-27 & AC-DC & $M_3$ & 1.6829 & 1.6741 & 1.6635 & & & \\ \hline

\multirow{9}{*}{\begin{tabular}[c]{@{}c@{}}32-bus\\ (30\% \\ DERs)\end{tabular}} & \multirow{9}{*}{12.66} & \multirow{9}{*}{17.90} & \multirow{3}{*}{0.0} & 6-7 & AC-DC & $M_1$ & 1.7481 & 1.7492 & 1.7350 & \multirow{3}{*}{5.847\%} & \multirow{3}{*}{5.847\%} & \multirow{3}{*}{5.957\%} \\ \cline{5-10}
 & & & & 12-13 & DC-AC & $M_2$ & 0.6175 & 0.6183 & 0.6215 & & & \\ \cline{5-10}
 & & & & 26-27 & AC-DC & $M_3$ & 1.7499 & 1.7490 & 1.7371 & & & \\ \cline{4-13}
 & & & \multirow{3}{*}{0.5} & 6-7 & AC-DC & $M_1$ & 1.6560 & 1.6561 & 1.6599 & \multirow{3}{*}{5.961\%} & \multirow{3}{*}{5.961\%} & \multirow{3}{*}{6.097\%} \\ \cline{5-10}
 & & & & 12-13 & DC-AC & $M_2$ & 0.6196 & 0.6195 & 0.6192 & & & \\ \cline{5-10}
 & & & & 26-27 & AC-DC & $M_3$ & 1.6578 & 1.6581 & 1.6621 & & & \\ \cline{4-13}
 & & & \multirow{3}{*}{0.9} & 6-7 & AC-DC & $M_1$ & 1.6560 & 1.6560 & 1.6551 & \multirow{3}{*}{5.970\%} & \multirow{3}{*}{6.169\%} & \multirow{3}{*}{6.268\%} \\ \cline{5-10}
 & & & & 12-13 & DC-AC & $M_2$ & 0.61883 & 0.6188 & 0.6192 & & & \\ \cline{5-10}
 & & & & 26-27 & AC-DC & $M_3$ & 1.6648 & 1.6650 & 1.6588 & & & \\ \hline\hline
\end{tabular}
}
\end{table}

\subsection{ Test Case I} 
The 33-bus distribution network of \citet{article42} has been modified to a 32-bus distribution network (Test Case I) by replacing the substation bus with a generation system. This study constructs the AC-DC network by eliminating reactive components from the DC sections. The system includes two AC-DC converters (rectifiers) and one DC-AC converter (inverter). The modulation indices $M$ for different converters are denoted as follows: $M_1$ for the converter between buses 6 and 7, $M_2$ for buses 12 and 13, and $M_3$ for buses 26 and 27. The AC-DC converters are positioned at bus pairs 6-7 and 26-27, while a DC-AC converter is positioned between buses 12 and 13. Additionally, the network features complete DC line sections extending from Bus 27 to Bus 32 and from Bus 7 to Bus 12. \par
\begin{figure}[h!]
\centering
\includegraphics[trim=0mm 0mm 0mm 0mm,width=8.2cm,height=5.4cm]{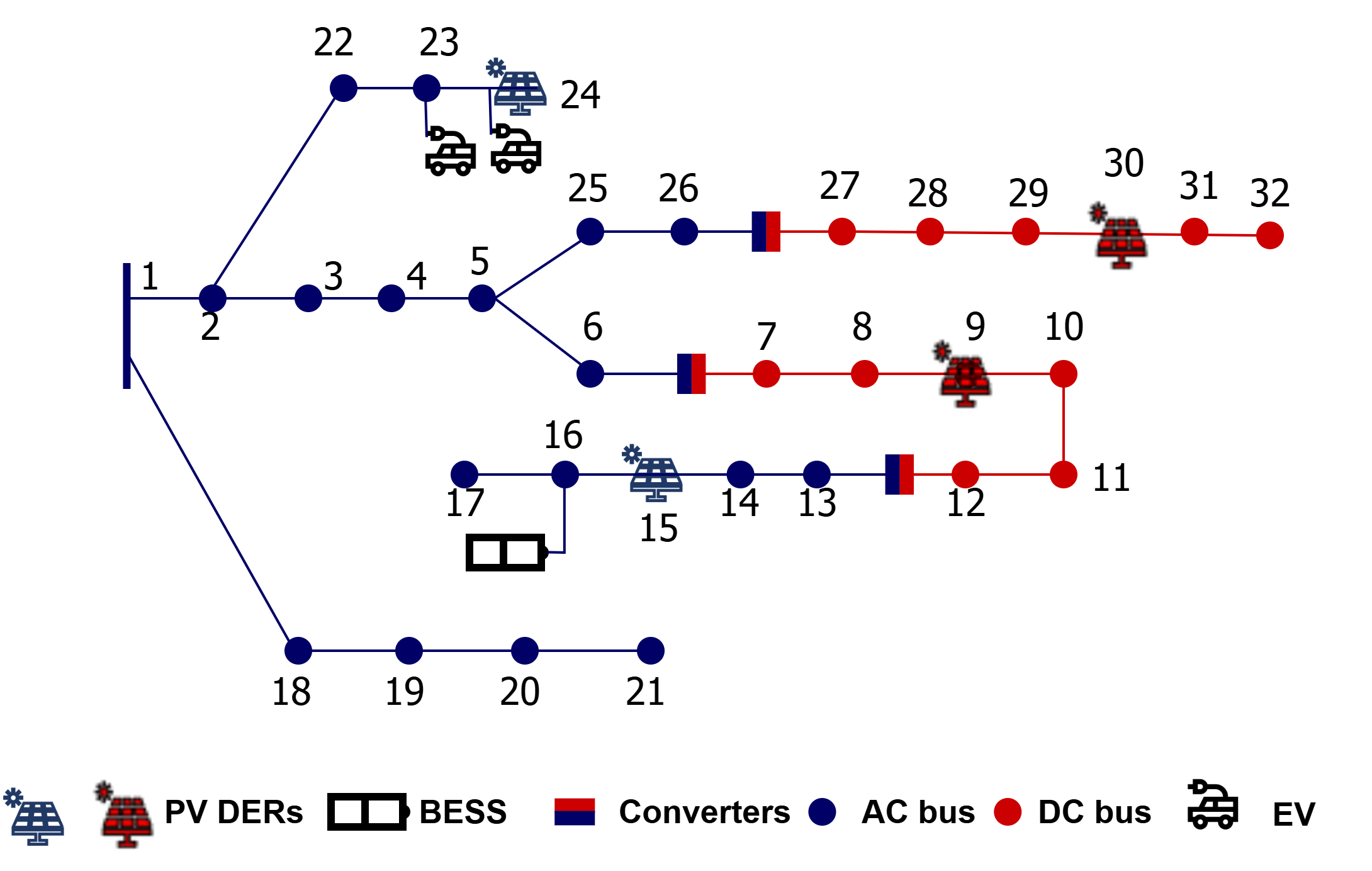}
\caption{Schematic diagram of the 32-bus AC-DC network with converters, EVs, BESS and PV DERs.}
\label{Fig: Figure3}
\end{figure}

The ratings of the three converters in sequence, as mentioned above, are $1200, 500$, and $1200$ KVA, respectively. The base voltage is set at 12.66 kV for the AC side and 17.90 kV for the DC side. The proposed OPF models are evaluated under the base case and a high DER integration scenario, where distributed generation (DGs) accounts for 30\% of the total connected load. The PV DERs are placed at bus $5,9,15,24$ and $30$. A BESS is placed at bus $16$. Two EV loads are considered for buses 23 and 24. The total load of buses 23 and 24 is $420+j200$ KVA each.  The maximum real power (kW) and reactive power (kVar) ratings of the PV DERs are 350, 100, 80, 350  \&  100 and 230, 0, 50, 230 \& 0, respectively. The capacity of the BESS is 80 kW. A schematic diagram of the network is illustrated in Fig.~\ref{Fig: Figure3}. \par

\begin{figure}[h!]
\centering
\includegraphics[trim=0mm 0mm 0mm 0mm,width=9cm,height=6.5cm]{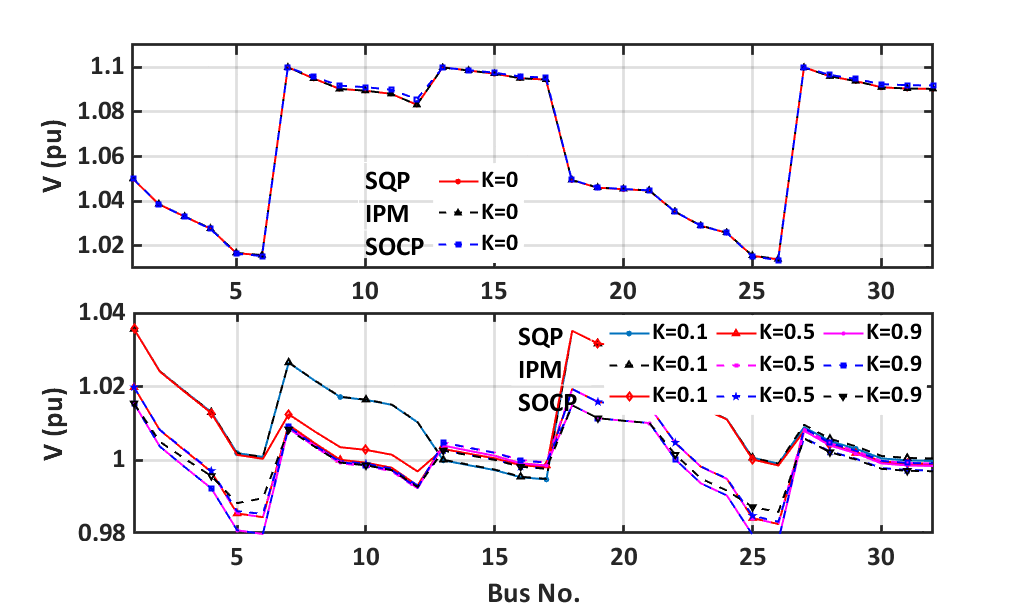}
\caption{Voltage profile with converters for the 32-bus network where legend SQP indicates the proposed NLP-SQP method while IPM and SOCP indicate IPM and SOCP-OPF method, respectively.}
\label{Fig:Figure4}
\end{figure}
The results are compared with those from the IPM-based method and SOCP-OPF results. In the experimental simulation, the converters operate with a maximum efficiency of 90\%, and the reference voltage is set to $V_i^{ref} = 1$ p.u. The steady-state voltage profile of the distribution network with integrated DERs is depicted in Fig. \ref{Fig:Figure4}. The voltage profiles obtained from all three OPF models exhibit similar trends, indicating consistent performance. A comparison of the optimal modulation index $(M)$ for the converters, as a function of the weight factor $(K)$, is presented in Table \ref{tab:Table IV}. \par
\begin{figure}[h!]
\centering
\includegraphics[trim=0mm 0mm 0mm 0mm,width=8.5cm,height=4cm]{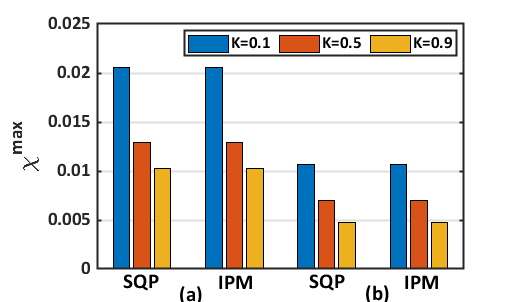}
\caption{Max. voltage deviation ($\chi^{max}$) at a bus. (a) 32-bus (30\% DERs), and (b) IEEE 123-bus (30\% DERs) networks.}
\label{Fig:Figure6}
\end{figure}

 
As the participation factor $K$ increases, the weighting of the objective function $f_2$ is enhanced, resulting in an improved voltage profile and a corresponding reduction in voltage deviation. The voltage mismatch is less than 0.01\%, and the modulation index $(M)$ for the proposed model is comparable with the IPM-based and SOCP-OPF models, validating the proposed NLP-SQP model. An increase in the weight factor $(K)$ leads to a reduction in voltage deviation from the reference voltage $(V_i^{ref})$, as illustrated in Fig. \ref{Fig:Figure6} in terms of $\chi^{max}$. The voltage deviation is computed as $|V_i - V_i^{ref}|$ for all buses in the network, where $|V_i - V_i^{ref}|$ denotes the maximum deviation across all buses. The convergence time of the OPF analysis for the 32-bus network is shown in Fig. \ref{bar_toc}. From the convergence time, it is seen that NLP-SQP takes less time than the IPM-based method because inequality constraints are hard to handle in the IPM algorithm for its combinatorial nature. In addition, the NLP-SQP's simulation time is close to the simulation time of SOCP-OPF.

\vspace{-3mm}
\subsection{Test Case II}
\begin{figure}[h!]
\centering
\includegraphics[trim=0mm 0mm 0mm 0mm,width=8.0cm,height=5.5cm]{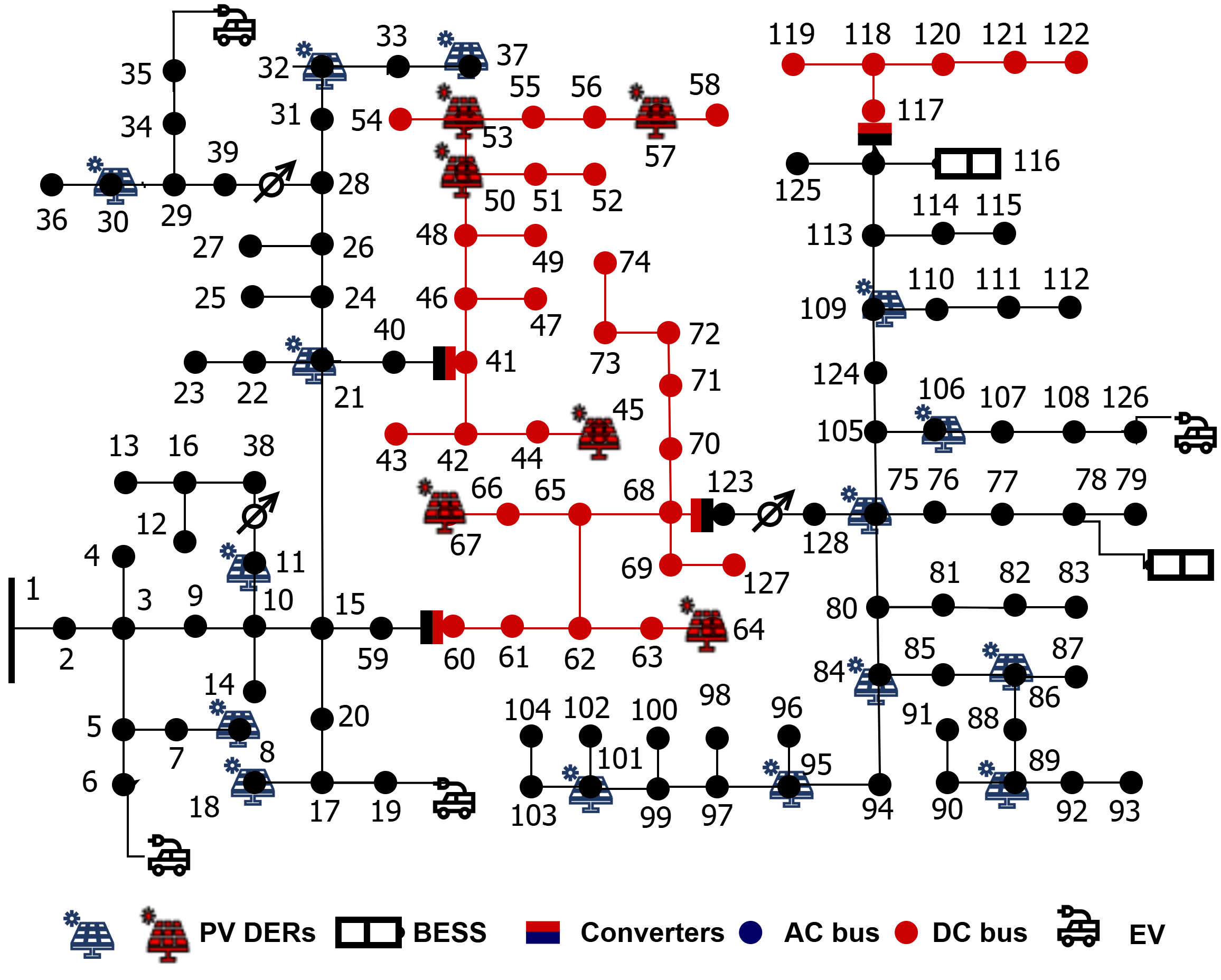}
\caption{Schematic diagram of the IEEE 123-bus AC-DC network (renumbered) with converters, BESS, EVs, and PV DERs.}
\label{Fig: Figure5}
\end{figure}

To assess the robustness and scalability of the proposed approach, simulations are performed on the IEEE 123-bus distribution network under two scenarios: the base case and a high DER integration case, where DERs account for 30\% of the total connected load. The conventional AC network is reconfigured into a hybrid AC-DC topology, as shown in Fig. \ref{Fig: Figure5}. Three AC-DC converters (rectifiers) and one DC-AC converter (inverter) are integrated into the network to form an AC–DC–AC architecture. The AC-DC converters, labeled $M_1$, $M_2$, and $M_4$, are connected between buses 40–41, 59–60, and 116–117, respectively. The DC-AC converter, $M_3$, connects buses 68 and 123. The corresponding converter ratings are 400 kVA, 1000 kVA, 100 kVA, and 800 kVA. Fig. \ref{Fig: Figure5} also indicates the locations of PV-based DERs, BESS, and EVs in the network. \par
\begin{figure}[h!]
\centering
\includegraphics[trim=0mm 0mm 0mm 0mm,width=8cm,height=4.8cm]{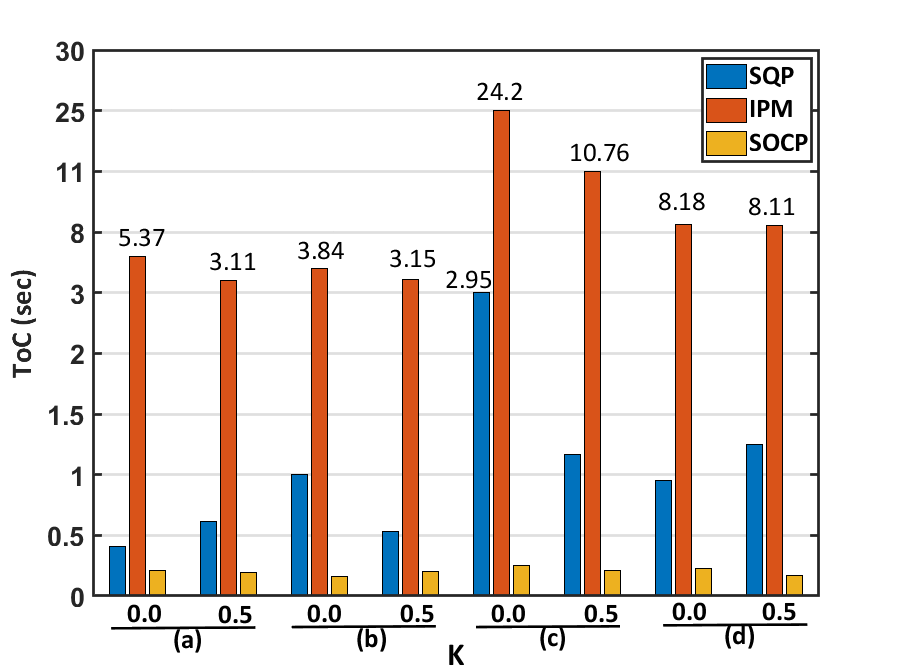}
\caption{Time of convergence comparison for NLP-SQP, IPM-based, and SOCP-OPF method for $K=0$ and $K=0.5$. (a) 32-bus network (base case), (b) 32-bus network (with 30\% DER); (c) IEEE 123-bus network (base case), and (d) IEEE 123-bus network (with 30\%DER).}
\label{bar_toc}
\end{figure}

The base voltage levels are set to 4.16 kV on the AC-AC side and 6.8 kV on the DC-DC side. All converters in the network are assumed to operate with a maximum efficiency of 90\%. The optimal operation point for the converters is shown in Table \ref{tab:Table V} for NLP-SQP, IPM-based, and SOCP-OPF methods for different weight factors $K$. As the weight factor $K$ increases, the total network loss slightly increases while the voltage deviation from the reference voltage decreases. Despite the marginal rise in losses, including the secondary objective function $f_2$ ensures voltage levels remain within stable and acceptable limits. The voltage mismatch among the evaluated OPF methods remains below 0.01\%, and the resulting modulation indices $(M)$ are closely aligned across all models, thereby validating the consistency and effectiveness of the proposed approach. $(\chi^{max})$ from $V_i^{ref}$ decreases with the increase of the weight factor $(K)$ as the 32-bus network, which is illustrated in Fig. \ref{Fig:Figure6}. The convergence time for Test Case II is given in Fig. \ref{bar_toc}, illustrating that the proposed NLP-SQP method is computationally comparable to the SOCP-OPF model.

\begin{table}[]
\centering
\caption{\textsc{Optimal 'M' and Percent of the Network Loss for the 123-bus Network }}
\label{tab:Table V}

\begin{tabular}{ccccccccc}
\hline\hline
\begin{tabular}[c]{@{}c@{}}Case\end{tabular} &
  K &
  \begin{tabular}[c]{@{}c@{}}Con.\\ ID\end{tabular} &
  \begin{tabular}[c]{@{}c@{}}Opt. \\ 'M'\\ SQP\end{tabular} &
  \begin{tabular}[c]{@{}c@{}}Opt. \\ 'M'\\ IPM\end{tabular} &
  \begin{tabular}[c]{@{}c@{}}Opt. \\ 'M'\\ SOCP\end{tabular} &
  \begin{tabular}[c]{@{}c@{}}Loss\\ (\%)\\ SQP\end{tabular} &
  \begin{tabular}[c]{@{}c@{}}Loss\\ (\%)\\ IPM\end{tabular} &
  \begin{tabular}[c]{@{}c@{}}Loss\\ (\%)\\ SOCP\end{tabular} \\ \hline
\multirow{12}{*}{\begin{tabular}[c]{@{}c@{}}Base\\ Case\end{tabular}} &
  \multirow{4}{*}{0} &
  $M_1$ &
  1.668 &
  1.667 &
  1.634 &
  \multirow{4}{*}{\begin{tabular}[c]{@{}c@{}}10.51\\ \%\end{tabular}} &
  \multirow{4}{*}{\begin{tabular}[c]{@{}c@{}}14.37\\ \%\end{tabular}} &
  \multirow{4}{*}{\begin{tabular}[c]{@{}c@{}}14.95\\ \%\end{tabular}} \\ \cline{3-6}
 &
   &
  $M_2$ &
  1.727 &
  1.755 &
  1.752 &
   &
   &
   \\ \cline{3-6}
 &
   &
  $M_3$ &
  0.571 &
  0.568 &
  0.570 &
   &
   &
   \\ \cline{3-6}
 &
   &
  $M_4$ &
  1.689 &
  1.692 &
  1.682 &
   &
   &
   \\ \cline{2-9} 
 &
  \multirow{4}{*}{0.5} &
  $M_1$ &
  1.639 &
  1.659 &
  1.642 &
  \multirow{4}{*}{\begin{tabular}[c]{@{}c@{}}11.20\\ \%\end{tabular}} &
  \multirow{4}{*}{\begin{tabular}[c]{@{}c@{}}14.69\\ \%\end{tabular}} &
  \multirow{4}{*}{\begin{tabular}[c]{@{}c@{}}15.34\\ \%\end{tabular}} \\ \cline{3-6}
 &
   &
  $M_2$ &
  1.660 &
  1.658 &
  1.652 &
   &
   &
   \\ \cline{3-6}
 &
   &
  $M_3$ &
  0.618 &
  0.614 &
  0.618 &
   &
   &
   \\ \cline{3-6}
 &
   &
  $M_4$ &
  1.639 &
  1.633 &
  1.638 &
   &
   &
   \\ \cline{2-9} 
 &
  \multirow{4}{*}{0.9} &
  $M_1$ &
  1.631 &
  1.631 &
  1.640 &
  \multirow{4}{*}{\begin{tabular}[c]{@{}c@{}}11.87\\ \%\end{tabular}} &
  \multirow{4}{*}{\begin{tabular}[c]{@{}c@{}}15.24\\ \%\end{tabular}} &
  \multirow{4}{*}{\begin{tabular}[c]{@{}c@{}}15.81\\ \%\end{tabular}} \\ \cline{3-6}
 &
   &
  $M_2$ &
  1.669 &
  1.669 &
  1.651 &
   &
   &
   \\ \cline{3-6}
 &
   &
  $M_3$ &
  0.619 &
  0.613 &
  0.618 &
   &
   &
   \\ \cline{3-6}
 &
   &
  $M_4$ &
  1.642 &
  1.623 &
  1.633 &
   &
   &
   \\ \hline
\multirow{12}{*}{\begin{tabular}[c]{@{}c@{}}30\%\\ DERs\end{tabular}} &
  \multirow{4}{*}{0} &
  $M_1$ &
  1.784 &
  1.703 &
  1.686 &
  \multirow{4}{*}{\begin{tabular}[c]{@{}c@{}}9.05\\ \%\end{tabular}} &
  \multirow{4}{*}{\begin{tabular}[c]{@{}c@{}}9.09\\ \%\end{tabular}} &
  \multirow{4}{*}{\begin{tabular}[c]{@{}c@{}}9.16\\ \%\end{tabular}} \\ \cline{3-6}
 &
   &
  $M_2$ &
  1.802 &
  1.765 &
  1.746 &
   &
   &
   \\ \cline{3-6}
 &
   &
  $M_3$ &
  0.564 &
  0.564 &
  0.566 &
   &
   &
   \\ \cline{3-6}
 &
   &
  $M_4$ &
  1.768 &
  1.765 &
  1.758 &
   &
   &
   \\ \cline{2-9} 
 &
  \multirow{4}{*}{0.5} &
  $M_1$ &
  1.638 &
  1.635 &
  1.642 &
  \multirow{4}{*}{\begin{tabular}[c]{@{}c@{}}9.16\\ \%\end{tabular}} &
  \multirow{4}{*}{\begin{tabular}[c]{@{}c@{}}9.23\\ \%\end{tabular}} &
  \multirow{4}{*}{\begin{tabular}[c]{@{}c@{}}9.34\\ \%\end{tabular}} \\ \cline{3-6}
 &
   &
  $M_2$ &
  1.641 &
  1.641 &
  1.648 &
   &
   &
   \\ \cline{3-6}
 &
   &
  $M_3$ &
  0.617 &
  0.610 &
  0.618 &
   &
   &
   \\ \cline{3-6}
 &
   &
  $M_4$ &
  1.639 &
  1.634 &
  1.639 &
   &
   &
   \\ \cline{2-9} 
 &
  \multirow{4}{*}{0.9} &
  $M_1$ &
  1.633 &
  1.633 &
  1.640 &
  \multirow{4}{*}{\begin{tabular}[c]{@{}c@{}}9.47\\ \%\end{tabular}} &
  \multirow{4}{*}{\begin{tabular}[c]{@{}c@{}}9.49\\ \%\end{tabular}} &
  \multirow{4}{*}{\begin{tabular}[c]{@{}c@{}}9.52\\ \%\end{tabular}} \\ \cline{3-6}
 &
   &
  $M_2$ &
  1.643 &
  1.642 &
  1.647 &
   &
   &
   \\ \cline{3-6}
 &
   &
  $M_3$ &
  0.614 &
  0.614 &
  0.618 &
   &
   &
   \\ \cline{3-6}
 &
   &
  $M_4$ &
  1.632 &
  1.632 &
  1.635 &
   &
   &
   \\ \hline\hline
\vspace{-4mm}
\end{tabular}

\end{table}

\begin{figure}[h!]
\centering
\includegraphics[trim=0mm 0mm 0mm 0mm,width=9cm,height=7.2cm]{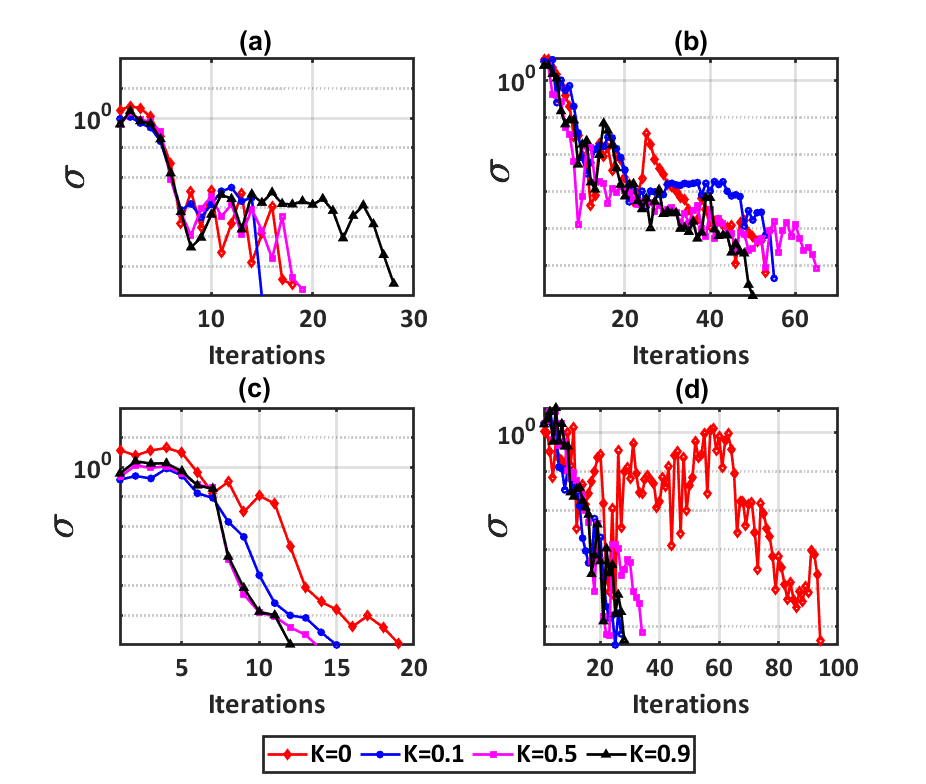}
\caption{Convergence analysis: 32-bus network (base case) (a) for NLP-SQP, (b) for NLP-based method; IEEE 123-bus network (with 30\%DER) (c) for NLP-SQP, and  (d) for IPM-based method.}
\vspace{-0.6cm}
\label{Figure_convergence}
\end{figure}

\subsection{Discussion on OPF Models}
In Fig.~\ref{Figure_convergence}, the convergence of NLP-SQP and IPM-based methods are illustrated for 32-bus modified distribution networks (base case) and the IEEE 123-bus distribution network (30\% DER case). The norm of the step length ($\sigma$) is plotted in Fig. \ref{Figure_convergence}. The figure shows that NLP-SQP and IPM-based methods converge for all values of $K$.  Moreover, it can be seen from the figure that for the same test case, NLP-SQP takes fewer iterations, and NLP-SQP converges faster. This is because the IPM-based method performs computationally expensive iterations as it moves towards gradient descent after performing each iteration in the presence of a high inequality-constrained optimization problem. However, after each iteration, NLP-SQP tries to guess which inequality constraints are active, updates the guess, and discards the inactive constraints. Similar results can be obtained from the 32-bus modified test case (with 30\%DER) and the IEEE 123-bus test case (base case). These results further prove the convergence of the proposed NLP-SQP method. On the other hand, though the SOCP-OPF method is faster than the NLP-SQP method, for reverse power flow, the solution from the SOCP-OPF method may not be exact always \citet{Gan_SOCP}. However, being convex, the SOCP-OPF solution is exact. So, the comparison with the solution from the SOCP-OPF method proves the global optimality of the proposed NLP-SQP method.    

\section{Conclusion }\label{conclusion}
This article presents and analyzes NLP-SQP methods for AC-DC radial power distribution networks. Unlike conventional approaches, the proposed methods optimize the operating points of distributed energy resources (DER) along with the optimal modulation index of the converters as a decision variable. Furthermore, the objective function explicitly accounts for voltage deviation caused by converter operation, thereby improving the network's voltage stability margin (VSM).\par

Extensive analysis on large-scale networks derived from standard distribution test systems demonstrates the superior performance of the NLP-SQP method. Compared with the IPM-based method, NLP-SQP exhibits faster convergence, greater scalability, and higher efficiency. Furthermore, it outperforms the SOCP-OPF method by achieving lower real power losses while maintaining an accurate and feasible solution. Given its ability to handle highly constrained AC-DC OPF problems effectively, the NLP-SQP method emerges as a robust and reliable optimization framework, and its scalability makes it well-suited for large-scale network analysis, reinforcing its potential for real-world applications.





 \bibliographystyle{elsarticle-harv} 
 \bibliography{reference.bib}





\end{document}